\documentstyle[aps,prb,epsf]{revtex}

\begin{document}

\twocolumn[\hsize\textwidth\columnwidth\hsize\csname
@twocolumnfalse\endcsname

\title{Low temperature properties of core-softened models: 
water vs. silica behavior}
\author{E. A. Jagla}
\address{Centro At\'omico Bariloche and Instituto Balseiro, 
Comisi\'on Nacional de Energ\'{\i}a
At\'omica \\(8400) S. C. de Bariloche, Argentina}
\maketitle
 
\begin{abstract}
A core-softened model of a glass forming fluid is numerically studied
in the limit of very low temperatures. The model shows two 
qualitatively different behaviors depending on the strength of
the attraction between particles. For no or low attraction, 
the changes of density as a function 
of pressure are smooth, although hysteretic due to mechanical metastabilities.
For larger attraction, sudden changes of density upon compressing and decompressing
occur. This global mechanical instability is correlated to the existence of a 
thermodynamic first-order amorphous-amorphous transition.
The two different behaviors obtained 
correspond qualitatively to the different phenomenology observed in silica and water.

\end{abstract}
 
\vskip2pc] \narrowtext

\section{Introduction}

The recently acknowledged possibility of the existence of single 
component systems which display coexistence between two different liquid phases 
has opened many interesting questions, and shed new light into the study of
the anomalous properties these systems display\cite{llequil1,llequil2,llequil3}. 
The case of water is probably
the most intensively studied, 
due to its ubiquity in nature. There 
is by now a general consensus that water displays a transition between
two different amorphous states in the supercooled region of its phase diagram
\cite{water}.
Experiments carried out in water at $T\simeq 130 K$ show an abrupt
change of volume $v$ as a function of pressure $P$,
which indicates the existence of the 
first order transition\cite{mishima}. The $v(P)$ curve
is hysteretic, and the jump in $v$ occurs
at $P\simeq 0.3$ GPa upon compressing, and at $P\simeq 0.05$ GPa 
upon decompressing. 
The volume change at the transition is about 0.2 cm$^3$/g. 
The two amorphous phases of water are 
smoothly related to two different liquid phases
at higher temperatures,
which at coexistence determine a first order liquid-liquid 
transition line ending in a critical point,
located in the metastable region of the phase diagram.
Most of the anomalies of water are usually interpreted as a 
consequence of the existence of this liquid-liquid
line and the liquid-liquid critical point, 
but in general the existence of liquid-liquid coexistence 
is not a necessary condition
for the existence of most of the anomalies\cite{jagla99}.

Water is not an isolated case. There is a whole family of substances,
usually referred to as tetrahedrally coordinated 
materials, that display many of the 
anomalies of water\cite{pccp}. Within this family,
another particularly interesting case is that of
amorphous SiO$_2$ (silica). 
Many of the anomalous properties of water are also found in silica.
Some of them have been observed in experiments (as the density anomaly) 
and other only in numerical simulations up to now (as maxima of 
isothermal compressibility and diffusivity as a function of pressure
\cite{poole}).
This has led to think that possibly a first order amorphous-amorphous
transition also occurs in silica.
But evidence of this transition has been elusive.
Experiments at ambient temperature in silica 
show an irreversible increase of density when the system is 
compressed up to $P\gtrsim 20$ GPa and successively decompressed\cite{siexp}. 
This behavior is reproduced in numerical simulations\cite{sinum}.
However, for no particular
value of pressure there is a sudden change in density that could be interpreted
as a direct evidence of a first order transition.

It has been argued that the qualitatively different behavior of water and silica 
is due to the temperature at which experiments are carried out compared
to the glass temperature $T_g$ of the materials\cite{ttg,poole,lacks}. Whereas $T/T_g$ is about 0.1
for the experiments in silica, 
it is close to 1 for water. Some people have raised the expectation
that if compressing experiments were done in silica at temperatures 
near or above $T_g$, they would
reveal the first order transition, which is supposed to be hidden in the ambient
temperature experiments due to lack of thermodynamic equilibrium.

We will not study the actual behavior
of water and silica. Instead, we will study a very simple model of a glass former
at low temperatures, including the limiting $T=0$ case, in different regions
of the parameters to gain qualitative insight into the real
cases of water and silica.
The model consists of particles interacting through a potential with 
a hard core plus a soft repulsive shoulder. In addition, 
an attractive contribution to the interaction can be included. 
This kind of model is not unrealistic for the study of the properties of 
tetrahedrally coordinated materials, and in fact it was shown to reproduce
many of the anomalies these materials possess\cite{jagla99,jagla00}.
We will show that when there is no attraction between particles (or only a weak one),
implying in particular that there is no first order transition, the $v(P)$
curve of our model in simulations at $T=0$
shows an hysteretic behavior, which is associated to the existence of different
mechanically stable configurations, and  qualitatively agrees with
the known phenomenology of silica. The inclusion of a sufficiently strong
attraction may produce the appearance of a first order transition, which is
clearly observed even in simulations at $T=0$, in the form of a global mechanical 
instability. The form of $v(P)$ we obtain in this case is very similar to what 
is found in water.

These findings will lead us to propose a 
different scenario to place together the properties of water and silica. 
We suggest that
the first order transition in water will be observable in the
form of a mechanical instability even in experiments at $T\rightarrow 0$. 
On the other hand we stress the possibility that 
there is no first order transition in silica, and that experiments performed in silica near $T_g$
will not reveal new qualitative ingredients. It is worth noticing 
that the existence of the thermodynamic anomalies of silica does 
not contradict this interpretation, since a liquid-liquid
critical point is not necessary to observe those anomalies, 
as it is known from studies in models 
closely related to the present one\cite{jagla99}.

The paper is organized as follows. The model and details on the simulation
procedure are presented in Section II. The results are contained in Section III, and
the relevance of them to silica and water is presented in
Section IV. Section V contains the summary and some final comments.

\section{The model}

The model we study is defined as follows.
We consider a bidisperse set of spherical particles, 
in order to avoid crystallization.
Particle $i$ is characterized by the value of a parameter $r_i$, which 
is taken from a bimodal distribution,
i.e., $r_i=r_a$, or $r_i=r_b$, with equal probability. 
The interaction potential $U$ between
particles $i$ and $j$ depends on $\tilde r\equiv r/R$ where $r$
is the real distance between particles, and $R=r_i+r_j$. The potential $U$ is composed
of a repulsive and an attractive part, $U\equiv U^R+U^A$.
The form we use for $U^R$ is
\begin{eqnarray}
U^R&=& \infty ~~~~~~~~~~~~~~~~{\rm for}~~~~ \tilde r<1\nonumber\\
U^R&=& \varepsilon_0 \frac{R}{\bar r}\left(\frac{0.01}{\tilde r-1} +1.2
-1.8(\tilde r-1.1)^2 \right)^2~~{\rm for}~~~~
1<\tilde r<1.9202\nonumber\\
U^R&=&0 ~~~~{\rm for}~~~~~ \tilde r>1.9202.  \label{vder}
\end{eqnarray}
where $\bar r=r_a+r_b$.
$U^R$ is plotted in Fig. \ref{potencial}. 
The form of this potential is a smooth version of a potential that we have studied
in detail previously\cite{jagla99,jagla98}. This smoother 
form is preferred here in order to avoid ambiguities in the calculations
at $T=0$, that appear in case the forces are not continuous.
The particular analytical form we use is not
really important, the only crucial feature of the potential 
is the existence of two preferred
distances between particles ($r_0\sim 1.1~R$ and $r_1\sim 1.9~R$ in Fig. 
\ref{potencial}) 
depending on the value of the external pressure.
The attractive part of the potential $U^A$ is simply given by
\begin{eqnarray}
U^A&=&\varepsilon_0\frac{R}{\bar r} 
\alpha(\tilde r- b)~~~~~~~~~~{\rm for}~~~~\tilde r<b
\nonumber\\
U^A&=&0~~~~~~~~~~~~~~~~~~~~~~{\rm for}~~~~\tilde r>b
\label{ua}
\end{eqnarray}
with the two dimensionless
parameters $b$ and $\alpha$ fixing respectively the range 
and intensity of the attraction. Two examples of the potential with the attraction
term (those to be used in the simulations) are also shown in Fig. \ref{potencial}.
In all the results to be presented (corresponding to two-dimensional
systems), we
use $r_a=0.45 \bar r$, and $r_b=0.55 \bar r$.
Temperature will be measured in units of $\varepsilon_0 k_B^{-1}$, 
pressure in units of $\varepsilon_0 k_B^{-1}\bar r ^{-2}$, and volume in
units of $\bar r^2$.

We simulate the system by standard Monte Carlo techniques. Particles are placed in a box with 
periodic boundary conditions.
At each time step the position of a single particle is modified to a new
position which is randomly chosen within a sphere of radius 0.01 
$\bar r$ centered at the original position. This trial movement 
is accepted according to the Metropolis rule. The update is made 
sequentially for all particles. 
Results using constant pressure simulations 
and others at constant volume will be shown. 
In the constant pressure scheme,
the volume of the system is
considered as an additional Monte Carlo variable, and
homogeneous expansion and contraction of the coordinates of all particles (and also
of the size of the simulation box) are tried.
This permits the volume of the system to adjust to the given external pressure.
In constant volume simulations, pressure is calculated as minus the energy change 
divided the volume change in a small (virtual) homogeneous rescaling of all coordinates of 
the particles.


\section{Results}

We will show results for a two dimensional system. This is done
to allow a better visualization of the configurations, 
however we emphasize that
all the results we discuss were qualitatively  
re-obtained in three dimensional systems.
We start by showing results in the case of no attraction 
($\alpha=0$), at $T=0$. 
The particles are randomly placed in space at the beginning of the simulation, 
and the system is rapidly compressed
up to reaching a mechanically stable configuration at $P\sim 0.5$, $v\sim 3.2$.
Then $P$ is increased, or $v$ is decreased 
(depending on the kind of simulations) by small steps. Mechanical equilibrium is obtained
at each $P$-$v$ value. In this way we reach $P\sim 4$, $v\sim 1.5$.
Then we slowly expanded the system to the original values of $P$ and $v$.

We see in Fig. \ref{t01} the values of $P$ and $v$ during this process, both
in simulations at constant $P$ an at constant $v$ for a system of 200 particles.
The curves show a series of small though 
abrupt changes in $v$ or $P$ (depending on the kind
of simulations) that correspond to mechanical instabilities 
in the system.
Also the hysteresis in $v(P)$ we observe
(which we found is repetitive upon 
compressing and decompressing, though only one cycle is shown in Fig.
\ref{t01}) originates in the existence of mechanical
metastabilities in the system. 
An indication of this fact
can be obtained from the following arguments. 
Let us define  $\tilde v\equiv\partial h/\partial P$, where $h=e+Pv$ is the
enthalpy per particle of the system as obtained from the simulations.
If the system was in thermodynamic equilibrium we should obtain
$\tilde v=v$, since this is a thermodynamic identity at $T=0$. We plot 
$\tilde v$ (from
the constant $P$ simulations)
also in Fig \ref{t01}.
We see that there are systematic differences between $v$ and
$\tilde v$, particularly in the region where $v$ changes rapidly. 
The difference between $v$  and $\tilde v$ is due to 
mechanical instabilities upon compressing and decompressing that produce
energy to be dissipated in the process. 
A very simple example is illuminating in this respect. 
For two particles interacting with the 
potential of Fig. \ref{potencial} (continuous line) at $T=0$, 
the evolution of distance $d$ (that replaces volume in this case) and 
enthalpy $h$ as a function of the compressing force $F$ (that takes 
the role of pressure) is shown in Fig. \ref{2p}.
The mechanical hysteresis in $d(F)$ is precisely the 
same effect we observe in the simulations, the only difference is that 
in the simulations the disordered nature of the system produces a smoothing of
$v(P)$.
In the two particle problem  $\tilde d\equiv
\partial h/\partial F$ equates $d$
except by the existence of two delta peaks at the points 
where there is a jump in $h$. This is the 
energy being dissipated. In the numerical 
simulations there is an averaging over many different atomic environments,
and the delta peaks are smoothed, but still clearly visible as regions
where $\tilde v<v$ during compressing, or $\tilde v>v$ during decompressing 
in Fig. \ref{t01}.

The mechanical instabilities in the system are of a local nature, in the
sense that they produce non-correlated rearrangements of particles as pressure
or volume changes. 
Then we expect the small reentrances in the constant $v$ simulations that
signal the existence of this instabilities to become weaker in larger samples, 
since they are averaged over the whole system. In fact, 
as an indication of this, we show in Fig. \ref{1000}
results for a system of 1000 particles, as compared to the case of
200 particles, and we see that `fluctuations' are considerably smaller. The global 
amount of hysteresis however remains quite the same.
Fig. \ref{snaps} shows snapshots of the 1000 particles system at the points indicated
in Fig. \ref{1000},and Fig. \ref{snaps2} shows the corresponding radial distribution 
function.
We see that as the pressure increases and volume decreases,
more and more particles move from a typical distance $r_1 \sim 1.9 R$ 
between neighbors to the shorter distance $r_0 \sim 1.1 R$. The collapse 
of neighbor particles from $r_1$ to $r_0$ is not a collective effect, 
it occurs in a non-correlated way in
different positions of the sample, namely no 
abrupt transition exists.

From the two particle problem we see an interesting characteristic of the $h(F)$
curve. The values of $h$ in the 
compressing and decompressing branches cross each other (at 
$F\bar r/\varepsilon_0 \sim 2.3$).
This crossing is also observed in the complete simulations. In fact, 
in Fig. \ref{h} we see values of $h$ obtained
during compression and decompression,
in the simulations at constant $P$ (all results are for a system of 1000
particles, from now on). We see clearly a crossing of the 
two branches at $P\simeq 1.8$. 
This crossing has no profound physical meaning, and in
particular it does not indicate the existence of a first order transition between
compressing and decompressing branches, since these
are not the thermodynamic values of the free energy, and the states accessed
during compression and decompression do not exhaust the whole phase space of
the system. To show this in more detail, 
we made simulations in which the system was annealed at fixed volume, 
trying to get as closer as possible to the real ground state of the system
at each $P$. 
The final values of $v$ that we get as $T\rightarrow 0$
at different $P$ are shown in Fig. \ref{equil}. 
Although we cannot 
guarantee that the true ground state
was obtained, the values of $h(P)$ in this case are systematically
lower than both the compression and decompression ones, 
as we can see in Fig \ref{h}. For the values of $v(P)$ obtained 
in the annealing simulations, the thermodynamic relation 
$v=\partial h/\partial P$ is well satisfied
within the numerical errors (Fig. \ref{equil}).
The values of $v(P)$ obtained in the annealing process
are inside the loop of compression-decompression, as it could be expected for the 
equilibrium values, since the hysteresis loop represents the maximum amount
of mechanical metastability that the system is able to sustain. In addition,
$v(P)$ is continuous, and this is consistent with the lack of a thermodynamic first 
order transition at $T=0$ in our model.

To obtain a thermodynamic first order transition, and to study the way in which it
reflects in the compression decompression results at $T=0$, a certain 
amount of attraction between particles must be included.
The existence of attraction in the system may have profound effects in its phase 
behavior. We will first discuss analytically the case in which an attraction 
energy $\gamma(r)<0$
of infinite range is added to the interparticle energy $U$.
In this case, the energy per particle $e$ gets an additional contribution
$\delta e $ of the form 
$\delta e=\int_V S(r)\gamma(r) d{\bf r}$,
$S(r)$ being the radial distribution function.
The assumed limit in which the range of $\gamma(r)$ goes to infinity, corresponds
to the case in which this integral is governed only by the limiting value of 
$S(r)$ as $r\rightarrow \infty$, i.e., it will depend only on the density of
the system. In this case, $S(r)$ 
itself is not affected by the attraction term. Then $\delta e$
takes the form $\delta e=-\tilde \gamma/v $ with some constant 
$\tilde \gamma >0$. This term is directly added to the free
energy $G$ of the system. When computing the equation of state from $\partial G/
\partial v=0$, the only difference with the case without 
attraction is that $P$ in the state equation is replaced by 
$P+\tilde \gamma/v^2$. Then from
the results of the simulations without attraction we can immediately get those for
a system with an attraction that has infinite range, just rescaling (self-consistently)
the pressure axis through $P\rightarrow P+\tilde \gamma/v^2$. 
The result of this procedure is shown in Fig. \ref{vdv}.
We see that due to the volume-dependent rescaling of $P$, and to the form of the
$v(P)$ curve with no attraction, a re-entrance in $v(P)$
may appear if the attraction is higher than some minimum. 
The reentrance in the thermodynamic $v(P)$ curve indicates the existence
of a first order transition between two amorphous phases of different 
densities. This reentrance appears also for the
compressing and decompressing branches, basically at the same value of $\tilde\gamma$. 
This means that in this case the mechanical instability at zero temperature exist if and only
if the system has a thermodynamic first order transition at $T=0$.
Then the global mechanical instability upon compression and decompression as pressure 
passes through the spinodal points at which $\partial v/\partial p$ 
becomes infinity (points C and C' in Fig. \ref{vdv}) is an indirect observation
of the first order transition. We emphasize that in the present case
the instability is global, in the sense that once it occurs, it involves a finite 
fraction of the whole system, contrary to instabilities in the case without attraction,
which are associated to individual particles.
We note that for a truly infinite range attraction (more precisely, if the
range of the attraction is much larger than the system size) the loop 
in $v(P)$ when there is a first order transition is physical, namely, 
it is observable in constant volume simulations, 
and no Maxwell construction can be invoked to flatten it out. In constant $P$
simulations instead, we would get an abrupt volume change when we go through the 
pressure corresponding to the spinodal points, and this is what we are referring to as a
global mechanical instability.
From Fig. \ref{vdv} we also see that the position 
of the first order loop moves towards lower pressures
as the strength of the attraction is increased, and then it can be completely
moved into the metastable $P<0$ region when the attraction is strong enough. 
Notice also that in cases in which the compressibility anomaly of the purely 
repulsive model (the rapid change in volume around $P\sim 2$ in our case) is weaker, 
it may happen that no first order transition appears 
at all, for any value of the attraction.

The previous analysis tells us that in the case of a long range attraction, the
existence of a first order thermodynamic transition and mechanical instabilities
in the compression-decompression path at $T=0$ are closely related, each of them
implying the existence of the other. We want to analyze now to what extent this
scenario can be extended to the case in which the attraction is short ranged.
Then we conducted  compressing-decompressing 
simulations using the finite range form (\ref{ua}) of 
the attraction.
We first use a rather large value $b=4$ for the attraction range, expecting
to reobtain basically the mean field phenomenology.
For weak attraction [$c\lesssim 0.1$ in (\ref{ua})], 
we get $v(P)$ curves which are slightly modified with respect
to the case of no attraction, but no qualitatively new results appear.
But if the attraction is strong enough, we obtain signs of global 
mechanical instabilities and a first order transition. 
The results of compression-decompression simulations are illustrated 
in Fig. \ref{atraccion} for $c=0.15$.
We see that constant $v$ simulations get a region (for $1.8<v<2.4$ 
upon compression) with small reentrances in the calculated values 
of $P$. In turn, in constant $P$ simulations we see an abrupt 
collapse of volume that jumps in a finite amount. This jump 
signals the occurrence of a global mechanical instability. 

The snapshots of the system when going through the coexistence 
region (Fig. \ref{bubbles})
show indeed that bubbles of the two 
amorphous phases coexist. 
Note the difference between the configurations in Fig. \ref{bubbles},
and those with no attraction 
during compression and decompression [Fig. \ref{snaps}(B) and (D)]. 
Here the particles that collapse
to the new phase tend to form well defined clusters in the sample, 
whereas in the other case the system remains uniform. This 
collective clustering is the
driving force of the mechanical instability.


In this case in which the attraction range is rather large we expect the
mean field arguments of the previous section to apply, and then that the
system has (in addition to the global mechanical instabilities at zero temperature)
a thermodynamic phase transition at sufficiently low temperatures. To
check directly for this transition is of course not an easy task in general, 
but for the current parameters it turns out that the transition is still
observable in equilibrium simulations at finite temperature, and then we
can be sure that it will persist down to zero temperature. In fact, 
to show only one indicator, 
the $v(P)$ function in constant $v$ simulations at $T=0.09$ is also 
plotted in Fig. \ref{atraccion}. We see that hysteresis upon 
compressing and decompressing 
has completely disappeared, as it should be since 
we are in thermodynamic equilibrium. The $v(P)$ curve has an abrupt change at 
$P\sim 0.75$, signaling the existence of a two phase coexistence region, and 
then a first order transition.

Next we simulated a case in which the attraction is much shorter ranged, specifically, we 
used the attraction form (\ref{ua}) with $\alpha=2.2$ and $c=1.0$ (the interaction 
potential in this case can be seen 
in Fig. \ref{potencial}). For these parameters we are still able to detect the
existence of the first order thermodynamic transition in simulations at finite temperature.
In Fig. \ref{corto} we plot the $v(P)$ curve at $T=0.09$, in which hysteresis
has disappeared, and signs of a first order transition at $P\sim 0.55$ are apparent. 
The compression-decompression curve at $T=0$
is shown also in Fig. \ref{corto}. We see that in this case in which the attraction 
is much shorter ranged, the signs of global mechanical instabilities 
are still clearly observable. Note also
that in this case, the decompression to $P=0$ 
(in the constant $P$ simulations) was such that the system was not able 
to regain its initial density, and remains in a densified structure.

In simulations performed with attraction of even shorter range, or 
when this is not strong enough, we did not observe global mechanical instabilities,
but neither a thermodynamic first order transition, at least in the temperature
ranges where equilibrium calculations can be done.

Let us briefly summarize the numerical results we have presented, 
before going to their relevance
to understand the phenomenology of water and silica. For the case of no 
attraction between
particles the $v(P)$ curve at zero temperature shows hysteresis upon 
compression-decompression, which is originated in local mechanical 
metastabilities. The thermodynamic $v(P)$ in this case is smooth 
and globally stable ($\partial v/\partial P <0$)
When an attraction of sufficiently long range is included, and 
in the case this is stronger than some minimum value we get both 
a thermodynamic phase transition and global mechanical instabilities 
at $T=0$. For attractions of shorter range, and in cases we can
guarantee the existence of a first order transition, we have 
observed that the global mechanical instability still exists.

\section{Unifying the silica and water phenomenology}

Silica and water display many thermodynamic anomalies. They include
the well known density anomaly (a temperature at which density is maximum), compressibility
and specific heat anomalies, 
and diffusivity anomalies.
These anomalies
do not require the existence of a first order liquid-liquid or amorphous-amorphous
transition to exist. In fact, basically all of them are found in the present
model with no attraction, i.e., when we can be sure that there is no first order 
transition\cite{jagla99}. 

Experimental and numerical evidence suggests that water possess 
a first order liquid-liquid equilibrium at low temperatures. 
Experimental evidence of a first order transition in silica is lacking. The
present model shows that hysteresis in $v(P)$ curves is by no means a strong
evidence of a first order transition if a global mechanical instability is 
not observed. And in fact, experiments show that silica is globally stable
upon compression and decompression, although displaying hysteretic behavior. 
Some numerical evidence of a first order transition in silica has been 
presented. In one case however \cite{lacks} the evidence was just the crossing 
of the free energies obtained during compression and decompression 
simulations at $T=0$ of model silica. This was
erroneously attributed to an underlying 
first order transition. As we showed in our model this
crossing occurs whether there is a first order transition or not, and it is due
to microscopic metastabilities. More
serious evidence come from the simulations by Voivod {\it et al.} 
\cite{sciortino}. They use
two different numerical models of silica and extrapolate high temperature
results to zero temperature, finding evidence of a first order transition.
Still, the results are preliminary, and based on extrapolations that call
for more detailed study.

We have shown that a very simple and transparent model has or has not
a first order transition depending on the strength of the attraction that is
included. We have shown simulations in which a mechanical instability at $T=0$ and a
thermodynamic first order transition exist.
The mechanical instability is qualitatively similar to that found
in water in experiments at $T\sim T_g$. This leads us to expect that water must
display the same instability even in experiments at much lower temperatures. 
To confirm this expectation it would be interesting if some of 
the models that are used to simulate water,
and that have shown liquid-liquid coexistence, were tried in the limit of zero
temperature to look for the mechanical instability, or even if compression decompression 
experiments were done in water at much lower temperatures.

In the case of a long range attraction between particles, 
we have shown that a thermodynamical
first order transition and global mechanical instabilities are closely correlated.
It is tempting to expect that for short range attraction this correlation 
remains, but this is not guaranteed.
In cases in which attraction is short ranged, and 
when the first order thermodynamical
transition is observable in equilibrium simulations, we have still 
observed global mechanical instabilities.
We have not been able to identify a set of 
parameters for our model where a thermodynamical transition occurs
in the lack of global mechanical instabilities. But this might be due to our 
incapacity of detecting a thermodynamic first order transition at very 
low temperatures when this transition does not show up at temperatures 
at which equilibrium
simulations are possible. 

On the basis of the present results we consider that the two possible scenarios
for silica are the following.
It may happen that silica does not have a first order transition. 
Our results for zero
or low intensity attraction correspond in fact to a 
case in which a first order transition does not exist, 
and the phenomenology we obtain
is closely related to that of silica\cite{nota}. Still, the other 
open possibility is in fact that silica has
a first order transition but this is not reflected in the zero temperature 
compression-decompression experiments.  
In this respect again, it would be nice if the
numerical models used for silica that seem to posses a first order 
transition (and in case in fact this is confirmed) 
were used at $T=0$ to search for a global mechanical instability.
If this is obtained, then those models will be shown to be no 
reliable to describe silica in this limit, 
since we know real silica does not have this instability. 
If the model does not show a mechanical instability instead, this would indicate
that the correlation between mechanical instabilities and a first order 
thermodynamical transition is not universal. 

\section{Summary and conclusions}

In this work we have studied a model of a glass forming fluid, consisting of 
spherical particles with a hard core, a repulsive shoulder and some amount 
of attraction. We investigated to what extent the existence of global 
mechanical instabilities at $T=0$ is correlated to the existence of a 
thermodynamical first order transition between two amorphous phases. We have 
obtained that the model without, or with weak attraction between particles
does not display neither a first order transition nor global mechanical 
instabilities. This behavior coincides with the known phenomenology of silica.
In the presence of a strong attraction between particles, both a first order 
amorphous-amorphous transition and global mechanical instabilities at $T=0$
were obtained. This scenario corresponds to the behavior of water.
Our work suggests that the difference between the phenomenology
of silica and water may be related to the lack of an amorphous-amorphous
transition in silica, in opposition to the existence of this transition in water.
This is not incompatible with the existence of coincident
thermodynamic anomalies in both cases.

\section{Acknowledgements}

This work was financially supported by Consejo Nacional de Investigaciones Cient\'{\i}ficas y
T\'ecnicas (CONICET), Argentina. Partial support from Fundaci\'on Antorchas is also 
acknowledged.

\begin{figure}
\narrowtext
\epsfxsize=3.3truein
\vbox{\hskip 0.05truein
\epsffile{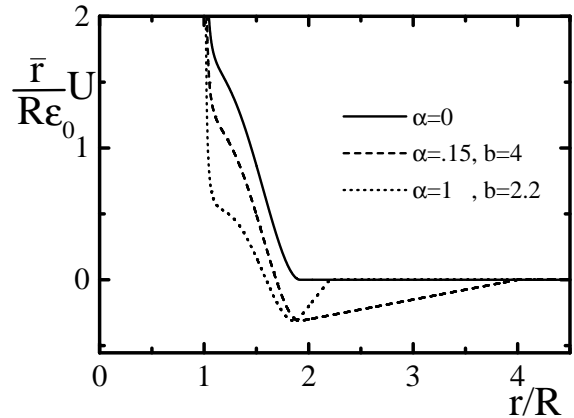}}
\caption{The interparticle potential $U(r)$ without attraction (continuous line) and
for two different choices of the attractive part.}
\label{potencial}
\end{figure}

\begin{figure}
\narrowtext
\epsfxsize=3.3truein
\vbox{\hskip 0.05truein
\epsffile{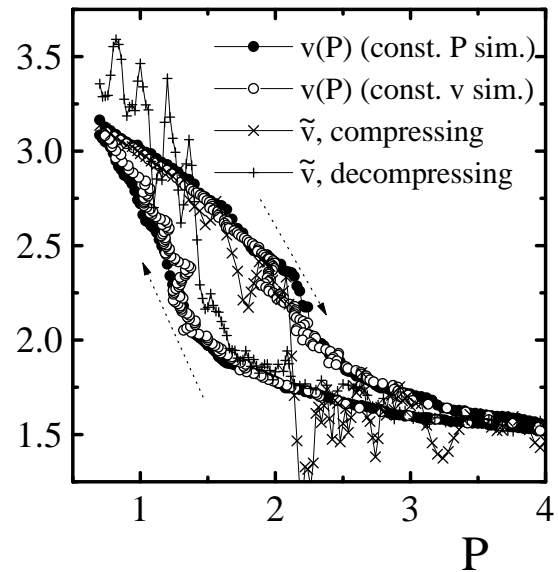}}
\caption{$v(P)$ function for the system without attraction, 
at $T=0$. Results of simulations at constant $P$ and
constant $v$ are shown. Also shown are the values of 
$\tilde v\equiv \partial h/\partial P$,
from the constant $P$ simulations. Differences between $v$ and $\tilde v$ 
reveal lack of thermodynamic equilibrium.}
\label{t01}
\end{figure}

\begin{figure}
\narrowtext
\epsfxsize=3.3truein
\vbox{\hskip 0.05truein
\epsffile{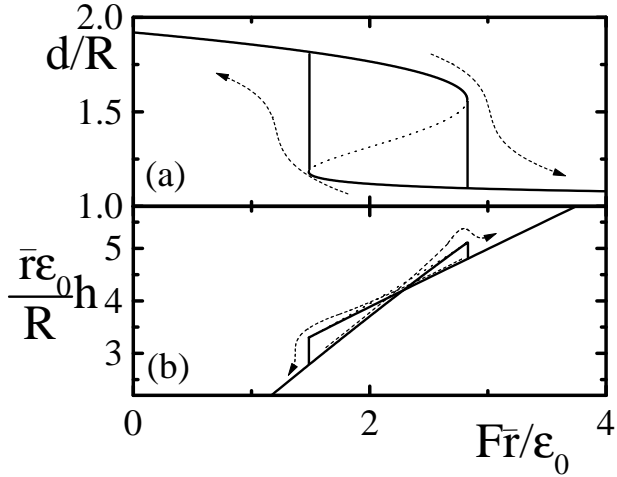}}
\caption{Distance $d$ (a) and enthalpy $h\equiv U+Fd$ (b) for two particles
interacting with the potential of Fig. {\protect\ref{potencial}} (continuous
line) as a function of the compressing force $F$. 
Compressing and decompressing routes are indicates.}
\label{2p}
\end{figure}

\begin{figure}
\narrowtext
\epsfxsize=3.3truein
\vbox{\hskip 0.05truein
\epsffile{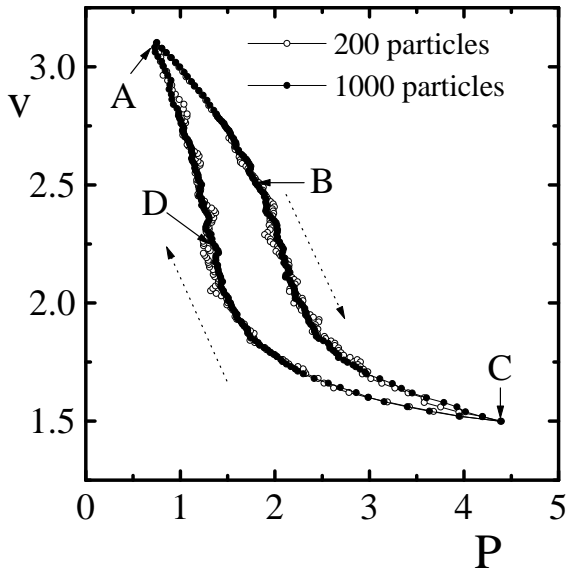}}
\caption{$v(P)$ curve as in Fig. \ref{t01}, for systems with 200 and 1000
particles. Fluctuations tend to average out in the larger sample, 
but the overall hysteresis remains the
same as in the case with fewer particles.} 
\label{1000}
\end{figure}

\begin{figure}
\narrowtext
\epsfxsize=3.3truein
\vbox{\hskip 0.05truein
\epsffile{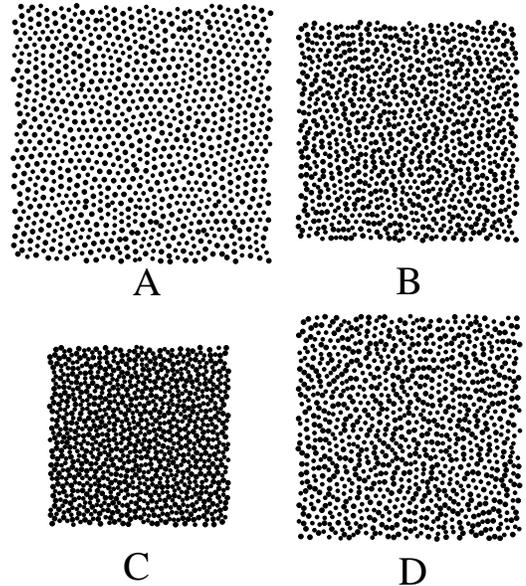}}
\caption{Snapshots of the system at the points indicated in Fig.
{\protect\ref{1000}}. Dot size represents the hard core of the particles.}
\label{snaps}
\end{figure}

\begin{figure}
\narrowtext
\epsfxsize=3.3truein
\vbox{\hskip 0.05truein
\epsffile{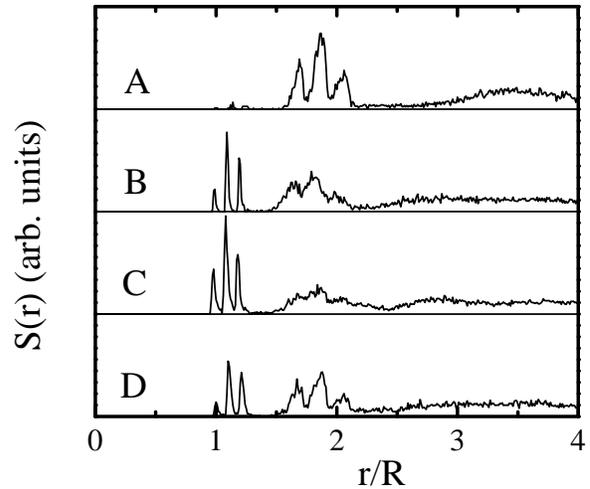}}
\caption{Radial distribution functions of the configurations shown in Fig. 
{\protect \ref{snaps}}. The weight transfer between $r_0\sim 1.1 R$ and $r_1\sim
1.9 R$ is clearly visible (the triplet structure of the peaks originates in the
bidispersity of the system).}
\label{snaps2}
\end{figure}

\begin{figure}
\narrowtext
\epsfxsize=3.3truein
\vbox{\hskip 0.05truein
\epsffile{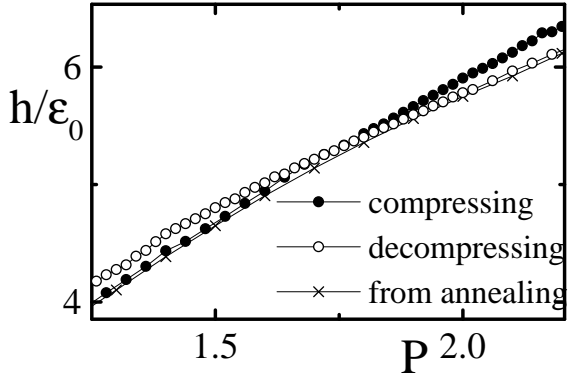}}
\caption{Evolution of enthalpy $h$ as a function of pressure during compression
and decompression, and from individual simulations annealing the system at 
fixed pressure. Compressing and decompressing branches cross each other at 
$P\sim 1.8$, but the thermodynamic values are always lower.}
\label{h}
\end{figure}

\begin{figure}
\narrowtext
\epsfxsize=3.3truein
\vbox{\hskip 0.05truein
\epsffile{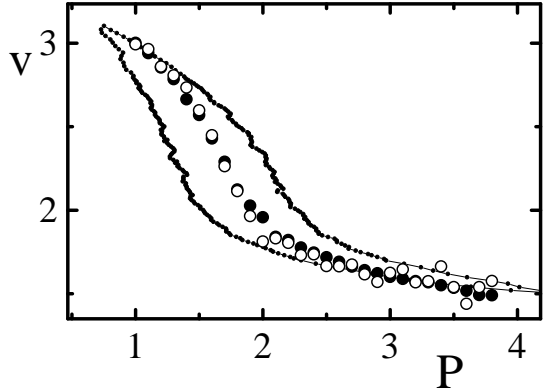}}
\caption{$v(p)$ curve obtained as the limit $T\rightarrow 0$ of individual 
simulations at constant volume (large full symbols), and calculated $\tilde v\equiv
\partial h/\partial P$ from the $h$ values obtained in the same simulations 
(open symbols). Within the numerical errors we get $v=\tilde v$, is it should be
in thermodynamic equilibrium. The compressing-decompressing loop of 
Fig. {\protect\ref{1000}} is also shown to allow comparison.}
\label{equil}
\end{figure}

\begin{figure}
\narrowtext
\epsfxsize=3.3truein
\vbox{\hskip 0.05truein
\epsffile{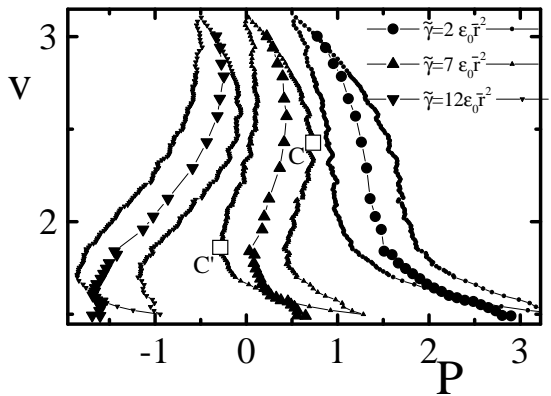}}
\caption{The effect of an infinite range attraction of different intensities
on the $v(P)$ curves of Fig. {\protect \ref{equil}}. A reentrance, indicative
of a first order transition 
is observed for the two largest values of $\tilde\gamma$.
For $\tilde \gamma \simeq 9\varepsilon_0\bar r^2$
the first
order transition is moved into the metastable $P<0$ region. 
Large symbols are the (almost) thermodynamic values, and small symbols are the
compressing and decompression results.
The position of the spinodal points $C$
and $C'$ upon compression and decompression is also shown in the case 
$\tilde \gamma=7\varepsilon_0\bar r^2$.}
\label{vdv}
\end{figure}

\begin{figure}
\narrowtext
\epsfxsize=3.3truein
\vbox{\hskip 0.05truein
\epsffile{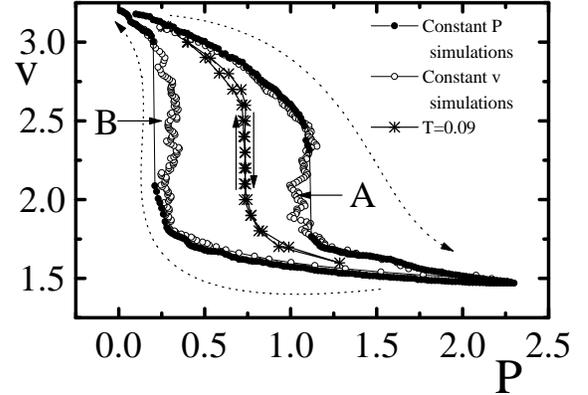}}
\caption{Full simulations with an attraction term of the form 
({\protect\ref{ua}}) with $b=4$, and $\alpha=0.15$. 
Constant pressure and constant volume simulations are shown.
Note the abrupt $v$ change in the constant $P$ simulations, 
and the weak reentrances of the constant $v$ simulations.
These are mechanical instabilities that indicate the existence of a thermodynamic
first order amorphous-amorphous transition in this case, which is made 
evident in the equilibrium $v(P)$ curve at $T=0.09$.}
\label{atraccion}
\end{figure}

\begin{figure}
\narrowtext
\epsfxsize=3.3truein
\vbox{\hskip 0.05truein
\epsffile{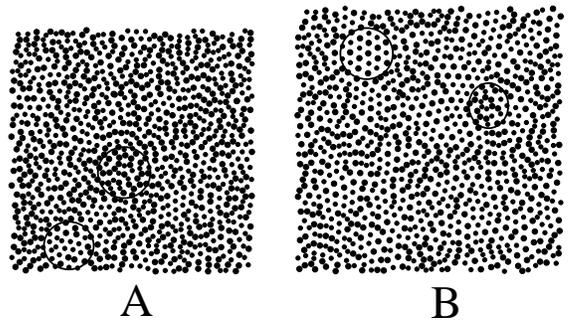}}
\caption{Snapshots of the system at the points indicated in Fig.
{\protect\ref{atraccion}}. Note the coexistence of rather large clusters
of two different phases with different densities, as for instance in the
encircled regions (compare with Fig. 
{\protect\ref{snaps}}).}
\label{bubbles}
\end{figure}

\begin{figure}
\narrowtext
\epsfxsize=3.3truein
\vbox{\hskip 0.05truein
\epsffile{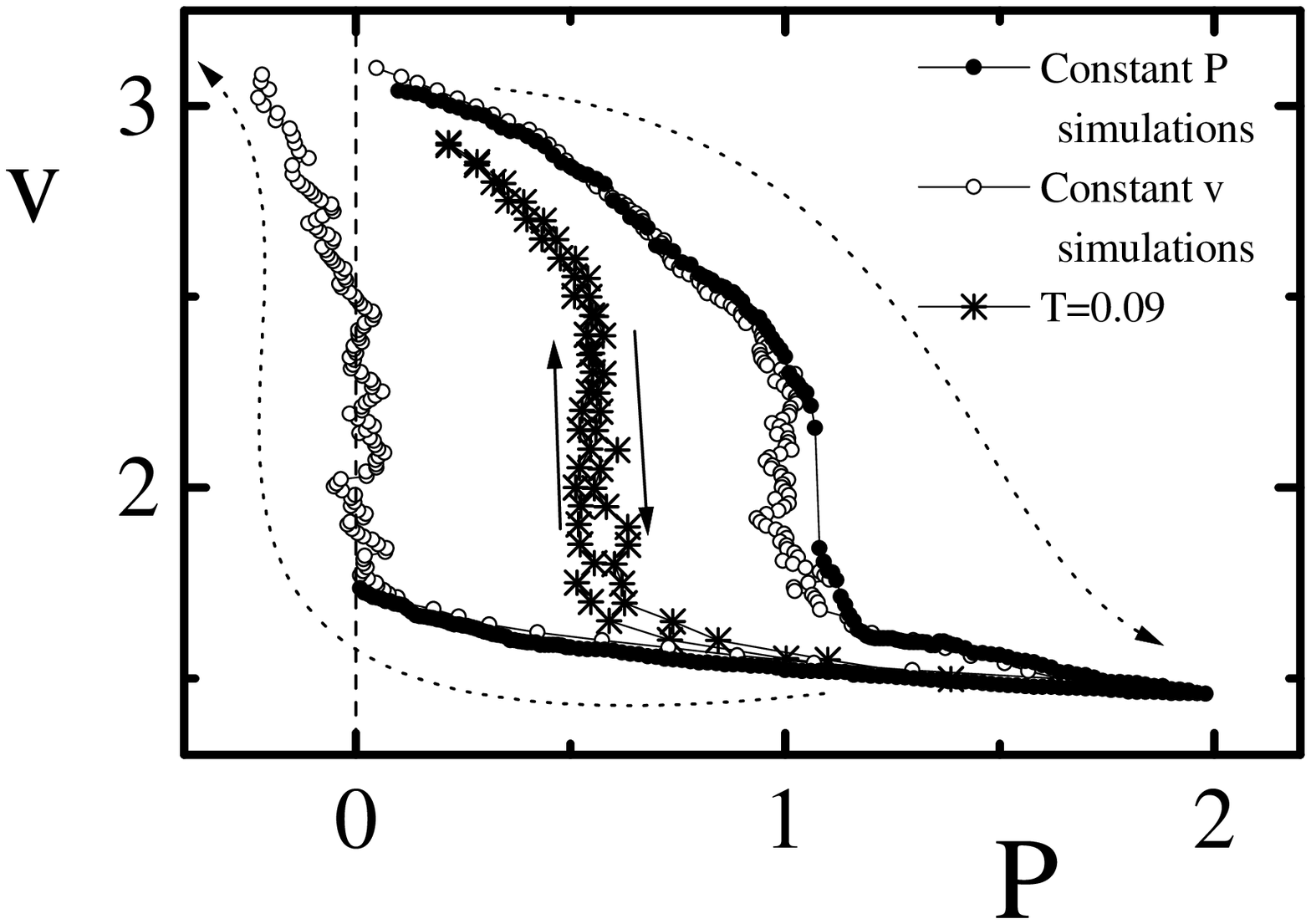}}
\caption{Same as Fig. \ref{atraccion}, but for $b=2$, and $\alpha=1.1$
(see Fig. \ref{potencial} for a sketch of the potential).}
\label{corto}
\end{figure}



\begin{references}

\bibitem{llequil1}P. H. Poole, F. Sciortino, U. Essman, and H. E. Stanley, Nature
{\bf 360}, 324 (1992).

\bibitem{llequil2}Y. Katayama, T. Mizutani, W. Utsumi, O. Shimomura,
M. Yamakaya, and K. Funakoshi, Nature {\bf 403}, 170 (2000).

\bibitem{llequil3}J. N. Glosli and F. H. Ree, Phys. Rev. Lett. {\bf 82}, 4659 (1999). 

\bibitem{water}O. Mishima and H. E. Stanley, Nature {\bf 396},
 329 (1998), and references therein.

\bibitem{mishima}O. Mishima, L. D. Calvert, and E. Whalley, Nature 
{\bf 310}, 393 (1984); O. Mishima, K.
Takemura, and K. Aoki, Science {\bf 254}, 406 (1991); O. Mishima, J. Chem.
Phys. {\bf 100}, 5910 (1994).

\bibitem{jagla99}E. A. Jagla, J. Chem. Phys. {\bf 111}, 8980 (1999).

\bibitem{pccp}C. A. Angell, R. D. Bressel, M. Hemmati, E. J. Sare, 
and J. C. Tucker, Phys. Chem. Chem. Phys. {\bf 2}, 1559 (2000).

\bibitem{poole}P. H. Poole, M. Hemati, and C. A. Angell, Phys. Rev. Lett.
{\bf 79}, 2281 (1997).

\bibitem{siexp}M. Grimsditch, Phys. Rev. Lett. {\bf 52}, 2379 (1984); 
R. J. Hemley, H. K. Mao, P. M. Bell, and B. O. Mysen,
Phys. Rev. Lett. {\bf 57} 747 (1986).

\bibitem{sinum}J. S. Tse, D. D. Klug, and Y. Lepage, Phys. Rev. B {\bf 46},
5933 (1992); W. Jin, R. K. Kalia, P. Vashishta, and J. P. Rino, Phys. Rev.
Lett. {\bf 71}, 3146 (1993); D. J. Lacks, Phys. Rev. Lett. {\bf 80}, 5385
(1998).

\bibitem{ttg}F. Sciortino, P. H. Poole, U. Essmann, and H. E. Stanley, 
Phys. Rev. E {\bf 55}, 727 (1997).

\bibitem{lacks}D. J. Lacks, Phys. Rev. Lett. {\bf 84}, 4629 (2000).

\bibitem{jagla00}E. A. Jagla, J. Phys.: Condens. Matt. {\bf 11}, 10251 (1999);
E. A. Jagla, Mol. Phys., to appear.

\bibitem{jagla98}E. A. Jagla, Phys. Rev. E {\bf 58}, 1478 (1998); 
J. Chem. Phys. {\bf 110}, 451 (1999).


\bibitem{sciortino}I. S. Voivod, F. Sciortino, and P. H. Poole,
preprint cond-mat/0007380.

\bibitem{jaglall}E. A. Jagla, preprint cond-mat/0006381.

\bibitem{nota} Experiments and simulations show that silica does not 
reach its original density upon compression-decompression, but remains
in a densified state, which is not exactly the behavior of our model without
attraction. But actually some amount of attraction exists indeed in silica,
and then this effect can be easily justified with our model, even in a case
in which a first order transition is absent.

\end{references}
\end{document}